\documentstyle[12pt]{ioplppt}

\def\lessim{\lower0.6ex\hbox{$\,$\vbox{\offinterlineskip
\hbox{$<$}\vskip1pt\hbox{$\sim$}}$\,$}}

\def\grtsim{\lower0.6ex\hbox{$\,$\vbox{\offinterlineskip
\hbox{$>$}\vskip1pt\hbox{$\sim$}}$\,$}}

\begin{document}

\title{TESTING THE EQUIVALENCE PRINCIPLE : WHY AND HOW
?\ftnote{7}{Talk given at the Symposium on Fundamental
Physics in Space (London, October 1995); submitted for
publication to Classical and Quantum Gravity.}}

\author{Thibault DAMOUR}

\address{Institut des Hautes Etudes Scientifiques, 91440
Bures-sur-Yvette, France}

\address{DARC-CNRS, Observatoire de Paris, 92195 Meudon, France}

\begin{abstract}
Part of the theoretical motivation for improving the present
level of testing of the equivalence principle is reviewed. The
general rationale for optimizing the choice of pairs of materials
to be tested is presented. One introduces a simplified rationale
based on a trichotomy of competing classes of
theoretical models. \end{abstract}

\section{Why testing the equivalence principle ?}

Einstein introduced in 1907 \cite{E7} what he called the
``hypothesis of complete physical equivalence'' between a
gravitational field and an accelerated system of reference. He
used this ``equivalence hypothesis'' \cite{E7}, \cite{E11} as a
heuristic tool to construct a physically satisfactory
relativistic theory of gravitation. The resulting theory,
general relativity, has been very successful, both in renewing
completely our description of the universe, and in passing with
flying colours all the experimental tests it has been submitted
to. For instance, the universality of free fall (the
experimental basis for the equivalence principle) has been
verified at the $10^{-12}$ level \cite{Su}, \cite{LLR}, the
quasi-static weak-field (``post-Newtonian'') predictions have
been checked in solar-system experiments at the $10^{-3}$ level
\cite{Solar}, the radiative structure (propagation of gravity
with the speed of light as an helicity-2 interaction) has been
verified by binary-pulsar data at the $10^{-3}$ level
\cite{Nobel}, and the quasi-static strong-field predictions have
been checked in binary-pulsar experiments at the $10^{-2}$ level
\cite{PSR}.

In view of this impressive record, should one apply Ockham's
razor and decide that Einstein's theory must be 100\% right, and
then stop testing it any further? My answer is definitely, no!
Indeed, one should continue testing a basic physical theory such
as general relativity to the utmost precision available simply
because it is one of the essential pillars of the framework of
physics. 

A less extreme attitude than Ockham's one might then be
to focus on the experimental tests which have presently the
lowest accuracy. In other words, this attitude would say that,
because $10^{-12} \ll 10^{-3}$, one should decide that Einstein's
equivalence principle is 100\% right, and concentrate on the
other tests of relativistic gravity. This point of view is the
one which has been traditionally taken by the American school of
``relativists'': Nordtvedt, Thorne, Will, $\ldots$, building on
foundations laid by Schiff and Dicke (see e.g. the book
\cite{W93}). I want, however, to emphasize that this bias
towards testing the class of so-called ``metric theories of
gravitation'', i.e. theories respecting the equivalence
principle, is quite unjustified, both from a historical
perspective, and (what is most important) from the point of view
of the current overall framework of fundamental physics.

First, from a historical perspective, the introduction of
well-motivated alternatives to general relativity is due to
Kaluza \cite{K21} and Jordan \cite{J+}. The motivation was that
a scalar field with gravitational-strength couplings appears
naturally as a new degree of freedom when one tries to unify
gravity with electromagnetism. A scalar field of the same type
(the ``dilaton'') appeared also naturally when Scherk and
Schwarz \cite{SS74} proposed the idea that string theory should
apply at the Planck scale. Still later, Scherk \cite{S79},
working within the framework of extended supergravities,
introduced the possibility of a vector field with
gravitational-strength couplings. From the point of view of the
present framework of physics, there are many reasons to expect
the existence of new interactions with strength related to the
gravitational one. In particular, extra vector fields appear
naturally in supersymmetry-inspired extensions of the standard
model \cite{F86}, \cite{Fayet}, and a plethora of (a priori)
massless scalar fields show up in string theories. Now, the main
point I wish to emphasize is that all the new interactions that
naturally appear in extensions of the present framework of
physics violate the equivalence principle. I know of no cases
where an exact ``metric'' coupling appeared naturally. The
historical reason why so much emphasis has been put in the
literature on artificially defined ``metric'' theories of
gravity comes from an important work of Fierz \cite{F56} on
Jordan's theory.

In this work, Fierz pointed out that Jordan's original theory
(with a Kaluza-Klein type scalar field) violated the equivalence
principle in an observationally unacceptable way. He then
introduced, in an {\it ad hoc} manner, the general class of
metrically-coupled tensor-scalar gravity theories (with one
arbitrary function) and the special one-parameter subclass of
Jordan-Fierz theories (often named after Brans and Dicke
\cite{BD61}).

The conclusion of all this is that the experiments which are the
most sensitive probes of new physics beyond the present
framework are tests of the equivalence principle. The fact that
present tests are at the $10^{-12}$ level does not diminish the
plausibility of small violations of the equivalence principle
because there exist string-inspired models \cite{DP94} in which
one gets, in a non fine-tuned way, violations of the
universality of free fall at the level
\begin{equation}
\frac{\Delta a}{a} \sim 10^{-18} \kappa^{-4} (\Delta
\varphi)^2  \, , \label{one}
\end{equation}
where $\kappa$ and $\Delta \varphi$ are dimensionless quantities
which could be of order unity. 

To illustrate the superior probing power of equivalence principle
tests let us mention that, barring the contrived possibility of a
``metric coupling'', there is always proportionality between
universality-of-free-fall deviations $\Delta a / a$ and
post-Newtonian deviations from general relativity (measured,
say, by the Eddington parameter $\overline{\gamma} \equiv
\gamma_{\rm Eddington} -1$). For instance, in the general class
of string-inspired models one can write \cite{DP94}, \cite{DVo95}
\begin{equation}
\left( \frac{\Delta a}{a} \right)_{AB} = \widehat{\delta}_A -
\widehat{\delta}_B  \, , \label{two}
\end{equation}
with
\begin{equation}
\widehat{\delta}_A = -\overline{\gamma} \left[ c_B
\left(\frac{B}{\mu} \right)_A + c_D \left( \frac{D}{\mu}
\right)_A + 0.943 \times 10^{-5} \left( \frac{E}{\mu}\right)_A
\right] \, . \label{three}
\end{equation}
Here, the suffixes $A,B$ label two material bodies whose free
falls are compared, while (in Eq. (\ref{three})), $\mu$ denotes
the mass in atomic mass units $B \equiv N+Z$ the baryon number,
$D=N-Z$ the neutron excess and $E=Z(Z-1)/(N+Z)^{1/3}$ a quantity
proportional to nuclear electrostatic energy. The third term on
the right-hand side of Eq. (\ref{three}) is expected to dominate
the other two. As the changes in $E/\mu$ can be $\grtsim 1$ (see
below) we see that, roughly speaking, $\Delta a / a \sim 10^{-5}
\overline{\gamma}$ in dilaton-like scalar models. In vector
models \cite{Fayet}, one has (from Eq. (3.13) of \cite{DEF1}
linking $\overline{\gamma}$ to the coupling of a new
interaction) $\Delta a /a \sim 10^{-2} \overline{\gamma}$ in the
generic case of a coupling significantly involving the lepton
number $L=Z$, and $\Delta a / a \sim 10^{-3} \overline{\gamma}$
in the particular case of a coupling only to $B=N+Z$. Therefore
equivalence principle tests constrain $\overline{\gamma}$ to the
$10^{-7}$ level in scalar models (see \cite{DVo95} for precise
numbers), and to the $\lessim 10^{-9}$ level in vector models.
The fact that this is much smaller than the $\vert
\overline{\gamma} \vert \lessim 10^{-3}$ level derived from
post-Newtonian or pulsar tests, gives a measure of the superior
probing power of equivalence principle tests.

\section{How to test the equivalence principle ?}

Taking the optimist view that improved equivalence principle
tests, henceforth abbreviated as EP tests, and notably STEP, will
give positive (i.e. non null) results, it is important to choose
the pairs of material tested so as to maximize, at once: (i) our
confidence in the reality of the EP violation signals, (ii) the
quantity of theoretical information that we can extract from the
experimental data. These questions have been addressed in
\cite{DB94} in some detail. We want here to summarize the main
points of \cite{DB94} and to propose a new, simplified approach
appropriate to possible descoped versions of STEP.

We assume that we are looking for EP violation signals caused by
some new long-range interaction. The interaction energy between
some laboratory body $A$ and an external body $E$ (the Earth in
STEP) reads
\begin{equation}
V_{AE} = -\frac{G m_A m_E}{r_{AE}} - \frac{H Q_A Q_E}{r_{AE}}
\equiv -G_{AE} \frac{m_A m_E}{r_{AE}} \, . \label{four}
\end{equation}
Here we have introduced an effective (composition-dependent)
gravitational constant for the $(AE)$ pair:
\begin{equation}
G_{AE} = G + H \frac{Q_A}{m_A} \frac{Q_E}{m_E} = G \left[ 1 +
\frac{H}{Gu^2} \frac{Q_A}{\mu_A} \frac{Q_E}{\mu_E} \right]  
\, , \label{five}
\end{equation}
where $G$ is Newton's bare gravitational constant, where $H$ is
the new coupling constant ($H>0$ for scalar exchange, and $H<0$
for vector exchange), and where $u$ denotes one atomic mass unit
so that $\mu_A \equiv m_A /u$. In the equations above $Q_A$
denotes the total ``charge'' of body $A$ to which the new
interaction is coupled.

The fractional difference in free fall acceleration of the pair
$(AB)$, $(\Delta a /a)_{AB} \equiv 2(a_A -a_B)/(a_A +a_B)$ is
given by
\begin{equation}
\left( \frac{\Delta a}{a} \right)_{AB} \simeq \frac{G_{AE}
-G_{BE}}{G} = \widehat{\delta}_A - \widehat{\delta}_B  \, ,
\label{six}
\end{equation}
with
\begin{equation}
\widehat{\delta}_A \equiv \frac{H}{Gu^2} \frac{Q_E}{\mu_E}
\frac{Q_A}{\mu_A} \equiv h \widehat{Q}_E \widehat{Q}_A \, .
\label{seven}
\end{equation}
Here we have defined the shorthands $h \equiv H/(Gu^2)$ and
$\widehat{Q}_A \equiv Q_A /\mu_A$.

The main issue of concern here is: how to optimize the choice of
materials to be tested? (see also \cite{Blaser}). As
emphasized in \cite{DB94}, if one does not assume any
theoretical model for the material dependence of the specific
charge $\widehat{Q}_A = Q_A / \mu_A$, the optimum strategy is:
(i) to restrict oneself to {\it connected} configurations of
test materials, i.e. set of pairs ${\cal C} = \{ (A_i A_j)\}$
such that any two elements $A_k$, $A_{\ell}$ can be connected by
a sequence of pairs belonging to ${\cal C}$, and (ii) to include
{\it topological loops} in the configuration ${\cal C}$ [e.g. a
null pair $(AA)$, a double pair $\{ (AB),(AB)\}$, a triangular
loop $\{ (AB), (BC), (CA)\}$, etc$\ldots$]. The reason for the
conclusion (i) is that the measurements of the left-hand sides
of Eq. (\ref{six}) determine only {\it differences} between
$\widehat{\delta}_A$'s, so that the choice of disconnected
configurations (e.g. $\{ (AB), (BC), (DE)\}$ introduces more
than one arbitrary additional constant in the phenomenological
determination of the $\widehat{\delta}_A$'s. The reason behind
the conclusion (ii) is that it allows one to exhibit convincing
checks of the reality of a violation of the EP which are
independent of any theory about the material dependence of
$Q_A$: e.g. the simple redundancy check $(\Delta a/a)_{(AB)_1} =
(\Delta a/a)_{(AB)_2}$ when using two (different in some respect)
pairs $AB$, or a richer cyclic check $(\Delta a/a)_{AB} +
(\Delta a/a)_{BC} + (\Delta a/a)_{CA} =0$. In view of the
practical difficulties in realizing null pairs or cyclic
configurations (see \cite{Blaser}, \cite{Lockerbie},
\cite{Touboul}), the inclusion of binary loops (repeated $(AB)$
with some difference) appears as the simplest way of confirming
the reality of an EP violation in a theory-independent manner.

It should be noted that the model-independent approach just
sketched cannot (even if one includes in ${\cal C}$ the maximum
possible number of independent pairs, say 91 to cover the
periodic table) give access to the basic theoretical quantities
$H$ and $Q_A$ (or rather $HQ_E^2$ and $Q_A / Q_E$, when taking
into account the possibility of arbitrary rescalings $Q_A
\rightarrow \lambda Q_A$, $H\rightarrow \lambda^{-2} H$). At
best, one can determine the $\widehat{\delta}_A$'s modulo an
arbitrary (common) additive constant. This is not even enough
information to determine the sign of $H$, i.e. the spin of the
mediating field.

Let us now shift to a model-dependent approach, i.e. assume some
theoretical model predicting a composition dependence of the
specific\ftnote{7}{``Specific'' is used here in the sense of
``per unit (atomic) mass''.} charge $\widehat{Q}_A \equiv Q_A /
\mu_A$ of the form
\begin{equation}
\widehat{Q}_A = \beta_0 + \sum_{i=1}^{n} \beta_i \xi_A^i \, .
\label{eight}
\end{equation}
Here $\beta_0$, $\beta_i$ are some coupling parameters and
$\xi_A^i$ some specific elementary charges. For instance, in
vector models (i.e. models where the apparent EP violation is
due to the exchange of an extra $U(1)$ long-range gauge field),
we expect to have only two independent elementary charges
\cite{Fayet} (for neutral bodies), baryon number and lepton
number, and no composition-independent coupling to mass: i.e. we
expect $\beta_0 = 0$, and $n=2$ with, say, $\xi^1 = (N+Z)/\mu$
and $\xi^2 = (N-Z)/\mu$. [Here, as above, $N=$ neutron number,
$Z=$ proton number $=$ atomic number $=$ lepton number.] On the
other hand, in the case of an EP violation due to any of the
long-range gauge-neutral scalar fields of string theory
(moduli), we expect to have a universal piece in $Q_A$, i.e.
$\beta_0 \not = 0$, and three independent elementary charges
\cite{DP94}. This yields for Eq. (\ref{eight}): $n=3$ and
\numparts
\begin{eqnarray}
\xi^1 &= &(N+Z)/\mu \, ,\hfill\label{ninea}\\
\xi^2 &= &(N-Z)/\mu \, ,\hfill\label{nineb}\\
\xi^3 &= &E/\mu \simeq Z(Z-1)/((N+Z)^{1/3} \mu) \, .\label{ninec}
\end{eqnarray}
\endnumparts
In Eq. (\ref{ninec}) $E$ denotes a contribution proportional to
the Coulomb interaction energy of a nucleus. A selection of the
values of the specific elementary charges (9) is presented
in table 1 (which is adapted from \cite{DB94}).

\bigskip

\begin{table}
\caption{
A selection of possible proof mass materials and their
corresponding specific elementary charges. Neutron numbers and
masses are averages weighted with natural isotope abundances.
}
\begin{indented}
\item[]\begin{tabular}{@{}llllll}
\br
Element &$Z$ &$N$ &$\left(\frac{N+Z}{\mu}-1\right) 10^3$ 
&$\frac{N-Z}{\mu}$ &$\frac{Z(Z-1)}{(N+Z)^{1/3}\mu}$\\
\mr
Be&{\lineup\0}4.&{\lineup\0\0}5.&$-$1.35175&0.110961&0.640133\\
C&{\lineup\0}6.&\lineup{\0\0}6.011&$-$0.003072&0.000916&1.09064\\ 
Mg&12.&{\lineup\0}12.3202&{\lineup\m}0.62322&0.013174&1.87451\\ 
Al&13.&{\lineup\0}14.&{\lineup\m}0.684212&0.037062&1.92724\\ 
Si&14.&{\lineup\0}14.1087&{\lineup\m}0.825719&0.003870&2.13129\\ 
Ti&22.&{\lineup\0}25.93&{\lineup\m}1.0772&0.082083&2.65644\\ 
V&23.&{\lineup\0}27.9975&{\lineup\m}1.09987&0.098103&2.67853\\ 
Cu&29.&{\lineup\0}34.6166&{\lineup\m}1.11663&0.088387&3.20096\\ 
Ge&32.&{\lineup\0}40.71&{\lineup\m}1.07046&0.119919&3.27228\\ 
Zr&40.&{\lineup\0}51.3184&{\lineup\m}1.0387&0.124073&3.7975\\ 
Ag&47.&{\lineup\0}60.9632&{\lineup\m}0.881352&0.129447&4.20924\\ 
Sn&50.&{\lineup\0}68.8079&{\lineup\m}0.822075&0.158435&4.19819\\ 
Ba&56.&{\lineup\0}81.4216&{\lineup\m}0.689875&0.185118&4.3462\\
Ta&73.&108.&{\lineup\m}0.287415&0.193425&5.13502\\
W&74.&109.898&{\lineup\m}0.266057&0.195257&5.16696\\
Pt&78.&117.116&{\lineup\m}0.18295&0.200511&5.30813\\
Au&79.&118.&{\lineup\m}0.169856&0.198003&5.37659\\
Bi&83.&126.&{\lineup\m}0.093913&0.205761&5.48788\\
U&92.&146.&$-$0.213316&0.226842&5.67502\\ \br
\end{tabular}
\end{indented}
\end{table}

{}From Eqs. (\ref{seven}) and (\ref{eight}) we conclude
that the theoretically expected composition dependence of
$\widehat{\delta}_A$ is of the form
\begin{equation}
\widehat{\delta}_A = \alpha_0 + \sum_{i=1}^{n} \alpha_i \xi_A^i
\label{ten}
\end{equation}
where $\alpha_0 \equiv h\left( \beta_0 + \sum_j \beta_j \xi_E^j
\right) \beta_0$, $\alpha_i \equiv h \left( \beta_0 + \sum_j
\beta_j \xi_E^j \right) \beta_i$.

The $A$-independent contribution $\alpha_0$ in Eq. (\ref{ten})
is not accessible from the measurements of Eq. (\ref{six}). The
best we can hope for is to measure the $n$ effective coupling
parameters $\alpha_i$. Once the $\alpha_i$'s are known, it will
be possible to measure the fundamental coupling parameters $h$,
$\beta_0$, $\beta_i$ (modulo the rescaling freedom $\beta_0
\rightarrow \lambda \beta_0$, $\beta_i \rightarrow \lambda
\beta_i$, $h \rightarrow \lambda^{-2} h$) {\it if and only if}
one knows (or assumes) something about the relative value of
$\beta_0$ with respect to the $\beta_i$'s. Such a knowledge is
available both in vector models $(\beta_0 =0)$ and in
string-scalar ones (see Eq. (6.13) of the first reference in
\cite{DP94}). Note that the composition of the Earth enters only
by introducing an $A$-independent proportionality factor between
the $\alpha_i$'s and the $\beta_i$'s, and does not influence the
strategy of choice of the configuration ${\cal C}$. The
nonlinearity of the relation $\alpha_i = \alpha_i (\beta_j)$ can
be tackled after having extracted (by least-squares fit) the
$\alpha$'s from the raw measurements.

Given the form (\ref{ten}), what is the optimal choice of
materials? This question has been addressed in \cite{DB94}.
Assuming that the measurement of the various differential
accelerations $(\Delta a/a)_{AB}^{\rm measured} \equiv m_{AB}$
can be modelled has containing independent gaussian errors,
$m_{AB} = \widehat{\delta}_A - \widehat{\delta}_B + n_{AB}$ with
$\langle n_{AB} n_{A'B'} \rangle = \sigma_{AB}^2
\delta_{A'B'}^{AB}$, one defines the likelihood function of the
theory parameters $\alpha_i$ (given a data set on some
configuration ${\cal C}$ of pairs of materials)
\begin{equation}
\chi^2 (\alpha_i) = \sum_{AB \in {\cal C}} \sigma_{AB}^{-2} 
(\widehat{\delta}_A - \widehat{\delta}_B -m_{AB})^2 = \sum_{AB}
\sigma_{AB}^{-2} \left( \sum_i \alpha_i \xi_{AB}^i
-m_{AB}\right)^2 , \label{eleven}
\end{equation}
where $\xi_{AB}^i \equiv \xi_A^i - \xi_B^i$. The minimum of the
function $\chi^2 (\alpha_i)$ then defines the best-fit values of
the $\alpha_i$'s. These $\alpha_i^{\rm best \, fit}$ are random
variables (when the realization of the noise changes) with
average values the true values of the $\alpha_i$'s and
deviations $\overline{\alpha}_i = \alpha_i^{\rm best \, fit} -
\alpha_i^{\rm true}$ some zero-mean gaussian variables with
distribution function $\propto \exp [-\Delta \chi^2 /2]$ where
$\Delta \chi^2 = \chi^2 (\alpha_i) -\chi^2_{\min}$ can be
written as
\begin{equation}
\Delta \chi^2 = \sum_{AB \in {\cal C}} \sigma_{AB}^{-2} \left[
\sum_i \overline{\alpha}_i \xi_{AB}^i \right]^2 = \sum_i \sum_j
g_{\cal C}^{ij} \overline{\alpha}_i \overline{\alpha}_j \, ,
\label{twelve}
\end{equation}
where one has defined
\begin{equation}
g_{\cal C}^{ij} \equiv \sum_{AB\in {\cal C}} \sigma_{AB}^{-2}
\xi_{AB}^i \xi_{AB}^j \, . \label{thirteen}
\end{equation}
In geometrical terms, the choice of a configuration of pairs of
materials ${\cal C} = \{ (A_a B_b)\}$ defines a quadratic form,
i.e. a {\it metric} $g_{\cal C}^{ij}$, Eq. (\ref{thirteen}), in
the $n$-dimensional space of coupling parameters $\alpha_i$
($\alpha$-space). The (natural, gaussian) ellipsoids of errors
of the $\alpha_i$'s are centered around $\alpha_i^{\rm true}$
and are defined by the above quadratic form: $\sum_{ij}
g_{\cal C}^{ij} \overline{\alpha}_i \overline{\alpha}_j \leq
\Delta \chi^2 = {\rm const.}$, where the value of $\Delta
\chi^2$ depends both on $n$ and on the chosen level of
confidence (e.g. if $n=2$ the ellipsoids $\Delta \chi^2 = 2.3$
and $\Delta \chi^2 = 6.2$ correspond to 68\% and 95\% confidence
regions, respectively). Optimizing the choice of configuration
${\cal C}$ means choosing a set of ``connecting vectors''
$\xi_{AB}^i \equiv \xi_A^i - \xi_B^i$ in the $n$-dimensional
$\xi$-space of specific elementary charges such that the
corresponding quadratic form (\ref{thirteen}) defines the
smallest and ``roundest'' ellipsoids in the dual $\alpha$-space.
A general geometrical rule for achieving this is to choose the
connecting vectors $\xi_{AB}^i$ so as to span the largest and
least degenerate (i.e. as far as possible from
$(n-1)$-dimensional configurations) volume in $\xi$-space.
Minimal configurations are made of $n$ vectors. The volume of
the ellipsoid of errors is inversely proportional to the volume
$\varepsilon_{ij\ldots \ell} \xi_1^i \xi_2^j \ldots
\xi_n^{\ell}$ spanned by the $n$ vectors, and its shape is
determined by the shape of the vectorial configuration
$\vec{\xi_1}, \vec{\xi_2}, \ldots ,\vec{\xi_n}$. [Note that the
geometry of $\xi$-space is purely {\it affine}, i.e. does not
make use of the concepts of (euclidean) length or angle.] To
illustrate this geometrical approach, we represent in Fig. 1,
using table 1, the three-dimensional $\xi$-space defined by Eqs.
(9). See \cite{DB94} for an application to the choice of a
configuration of pairs of materials. 

The strategy just explained is appropriate to an ambitious
experiment (such as M3STEP) which aims, at once, to establish
convincingly the existence of an EP violation, and to maximize
the precision of the simultaneous measurement of the underlying
theoretical coupling parameters. Such an experiment requires a
minimum of four differential accelerometers: three of them
spanning as large a volume as possible in $\xi$-space, and the
fourth one providing a redundancy check (closing a polygon or
repeating an edge). One can, however, settle for less ambitious
strategies if one considers descoped versions of STEP. I wish now
to introduce such a strategy. The basic idea is to argue that
theoretical expectations suggest plausible relative
orders of magnitude for the coupling parameters $\beta_i$ (and
thereby $\alpha_i$), and thereby put constraints on the
composition-dependence of the $\widehat{\delta}_A$'s.

First, in vector models it is plausible that the couplings to
baryon number and lepton number be of the same order of
magnitude, i.e. $\beta_1 \sim \beta_2$ and therefore $\alpha_1
\sim \alpha_2$. An example of this is provided by grand unified
theories which predict \cite{Fayet} a coupling to $B-L = N
=\frac{1}{2} (N+Z) + \frac{1}{2} (N-Z)$. As, numerically, $\xi^1
=B/\mu$ varies much less over the periodic table than $\xi^2 =
(N-Z)/\mu$ (see Table I), we conclude that, in most cases, the
vector specific charge can be well approximated by $\xi^2$ or,
nearly equivalently, by $N/\mu$. We should, however, consider
also the possibility that the vector charge be {\it exactly}
proportional to baryon number. In other words, we argue that the
one-dimensional continuum of possible values of $\alpha_1 /
\alpha_2$ can be, in first approximation, reduced to a
dichotomy between a charge $\propto (B-L)$ and a charge
$\propto B$.

Second, in string-inspired scalar models one has put plausible
constraints on the magnitudes of $\beta_1$ and $\beta_2$ (see
\cite{DP94}, p.552) which are such that the
composition-dependence in Eq. (\ref{three}) is numerically
dominated by the last term, i.e. by the coupling to the nuclear
electrostatic energy. In other words, we argue that the
two-dimensional continuum of possible values of $\alpha_1
/\alpha_3$, $\alpha_2 / \alpha_3$ can be, in first approximation,
reduced to a unique coupling $\propto E$. We have checked
numerically the above assertions by plotting the variation over
the periodic table ($Z$-dependence) of $\widehat{\delta}_A$
(normalized by imposing $\widehat{\delta}_{\rm Be} = 0$ and 
$\widehat{\delta}_{\rm Au} = 1$) in various theoretical models:
couplings to $L$, $B-L$, $B\pm 3L$, $E$ and $E\pm 240B \pm 5.2L$
(the latter coefficients being the maximal ones suggested in
\cite{DP94}). Instead of spreading all over the plane, the
above curves gather themselves in two well-separated bundles of
curves centered around the curves defined by couplings to $L$
and $E$.

Our  conclusion is that, in first approximation (i.e. barring
values of the coupling parameters that do not appear {\it a
priori} justified within the framework of present theoretical
ideas), the search among the continuum of theoretical models can
be simplified into a search among only {\it three} basic
possibilities: a coupling to $L$ (or $B-L$), a coupling to $B$,
or a coupling to $E$. This trichotomy is illustrated in
Fig. 2 which presents the corresponding $Z$-dependences of the
EP violation signals $\widehat{\delta}_A$ (normalized by an
affine transformation $\widehat{\delta}_A^{\rm new} = a 
\widehat{\delta}_A^{\rm old} +b$ so that $\widehat{\delta}_{\rm
Be}^{\rm new} = 0$ and $\widehat{\delta}_{\rm Au}^{\rm new} =
1$). We conclude from Fig. 2 that, in order to distinguish between
theoretical models, it is very important to include among tested
materials: Beryllium, Platinum (or Gold) and an element among
$\{$C, Mg, Si, Al$\}$, say Silicon (maybe in the form of silica
SiO$_2$). [Magnesium is probably too reactive to be considered
seriously.] This leads to a minimum of {\it two} pairs connecting
the three elements Be, Pt, Si, to which must be added a third,
redundant pair (e.g. a repeated (Be Pt)) to include a
theory-independent check on the reality of EP violation.

In all the above strategies, one was trying to get some definite
information about the theoretical nature of an EP violation. If
one has to descope a space mission to the minimum meaningful
concept, one can go down to {\it two} pairs of materials
containing the same materials, say the pairs $(AB)$, $(AB)'$,
where the prime indicates that some difference (in shape,
mass,$\ldots$) is introduced. The choice $A= {\rm Be}$, $B= {\rm
Pt}$ seems appropriate for maximizing the expected signal in
most models (see Fig. 2). Such a trimmed concept is appropriate
to a discovery experiment where one puts the emphasis on
establishing (by a redundancy) the reality of an EP violation,
rather than on extracting theoretical information from the data.

\section{Conclusions}

The main points emphasized above can be summarized as follows
\begin{itemize}
\item In spite of its impressive name, and of its having been put
on a pedestal by part of the ``relativity literature'', the
``Equivalence Principle'' is not at all a basic taboo principle
of physics. On the contrary, nearly all the attempts to extend
the present framework of physics (Kaluza-Klein, strings,
supersymmetry,$\ldots$) predict the existence of new
interactions (mediated by scalar or vector fields) violating the
universality of free fall.
\item Though present tests of the universality of free fall are
at the $10^{-12}$ level, there exist string-inspired models
(containing no small parameters) in which the theoretically
expected level of EP violation is naturally $\ll 10^{-12}$.
Equivalence Principle tests are, by far, the most sensitive
low-energy probes of such new physics beyond the present
framework.
\item Present theoretical models suggest a rationale for
optimizing the choice of pairs of materials to be tested in EP
experiments. In descoped experiments, one can trim this general
rationale down to a simple trichotomy among competing
classes of theoretical models. This leads to comparing Be, Pt
and Si (or SiO$_2$, or C, or Al). In a discovery experiment, one
can use only two pairs $(AB)$, $(AB)'$ (with some difference).
\end{itemize}

\Figures
\begin{figure}
\caption{Position of the elements in the
three-dimensional $\xi$-space of specific elementary charges. The
$x$, $y$ and $z$ axes are proportional to $\xi^1 -1$, $\xi^2$ and
$\xi^3$ of Eqs. (9).}
\end{figure}

\begin{figure}
\caption{The observable (normalized) violation of the equivalence
principle $\widehat{\delta}_A^{\rm new}$ as a function of atomic
number $Z$ plotted for couplings to $B$ (upper curve), $E$
(intermediate curve) and $B-L$ (lower curve). The latter two
curves are typical representatives of large continua of
(respectively) vector and scalar models having coupling
parameters of a naturally expected order of magnitude.}
\end{figure}

\bigskip

\end{document}